\documentclass[preprint,12pt]{elsarticle}
\usepackage[T1]{fontenc}
\usepackage[utf8]{inputenc}
\usepackage{lmodern}
\usepackage{microtype}
\usepackage{amsmath,amssymb}
\usepackage{graphicx}
\usepackage{subcaption}
\usepackage{booktabs}
\usepackage{array}
\usepackage[table]{xcolor}
\usepackage{lineno}
\usepackage{url}
\usepackage[colorlinks=true,allcolors=blue]{hyperref}
\usepackage{enumitem}

\definecolor{confirmcolor}{RGB}{160,0,120}

\newcommand{\EMIBF}{EMI--BF$_4$}
\newcommand{\EMIplus}{EMI$^+$}
\newcommand{\BFminus}{BF$_4^-$}
\newcommand{\MACEPolar}{MACE-POLAR-1}
\newcommand{\MACEMedium}{MACE-MP-0 (medium)}
\newcommand{\angstrom}{\text{\AA}}
\newcommand{\eV}{\ensuremath{\mathrm{eV}}}

\journal{Acta Astronautica}

\newcommand{\CorrespondingAuthor}{Ziyu Huang}
\newcommand{\CorrespondingEmail}{zyuhuang@gatech.edu}

\hypersetup{
  pdftitle={Machine-learning interatomic potentials for ionic-liquid fragmentation in electrospray thrusters},
  pdfauthor={\CorrespondingAuthor},
  pdfsubject={Benchmarking MACE foundation models for EMI-BF4 surface-impact fragmentation},
  pdfkeywords={electrospray thruster, ionic liquid, machine-learning interatomic potential, fragmentation, molecular dynamics}
}

\begin{document}
\begin{frontmatter}

\title{Simulating Ionic Liquid Fragmentation in Electrospray Thrusters with Foundation Models}

\author[gatech]{\CorrespondingAuthor\corref{cor1}}
\ead{\CorrespondingEmail}
\cortext[cor1]{Corresponding author}
\affiliation[gatech]{organization={Daniel Guggenheim School of Aerospace Engineering, Georgia Institute of Technology},
            city={Atlanta},
            state={Georgia},
            postcode={30332},
            country={USA}}

\begin{abstract}
Predicting the products of ionic-liquid impacts on extractor surfaces is important for electrospray-thruster lifetime analysis, yet available atomistic methods require a compromise between chemical fidelity and computational cost. Reactive force fields enable high-throughput sampling but do not explicitly resolve electronic charge redistribution and may miss relevant reaction pathways during impact, whereas mixed quantum--classical density-functional-theory molecular dynamics (DFT/MD) can capture charge redistribution and neutral-product formation at substantially higher computational cost. Pretrained atomistic foundation models have recently emerged as a potential route toward DFT-like chemical fidelity at considerably lower cost. Here, we benchmark two pretrained machine-learning interatomic potentials, \MACEMedium\ and \MACEPolar, against DFT/MD and ReaxFF for geometry optimization of 1-ethyl-3-methylimidazolium tetrafluoroborate (\EMIBF) and for 10--100~\eV\ impacts on a model Au extractor surface. The models reproduce several collision outcomes observed in DFT/MD, including ionic dissociation, high-energy covalent fragmentation, and, in particular, HF formation through neutralization-like chemistry that is not captured in the ReaxFF simulations. In the computational-performance benchmark, \MACEPolar\ and \MACEMedium\ completed each 2~ps trajectory in 5.12 and 2.54~min, respectively, corresponding to wall times approximately four orders of magnitude shorter than the DFT/MD reference under the reported benchmark conditions. These results support pretrained machine-learning potentials as a practical intermediate-cost approach for chemically resolved electrospray-impact simulations and motivate targeted fine-tuning with DFT data for broader applications in electrospray-thruster and electric-propulsion modeling.

\end{abstract}


\end{frontmatter}


\section{Introduction}

Ionic-liquid electrospray thrusters combine compact hardware, high specific impulse, and precise thrust control, making them attractive for small-spacecraft propulsion and drag-free missions. Their usable lifetime, however, is not determined by ion emission alone. Off-axis ions, clusters, and neutrals can collide with extractor and accelerator grids, where they deposit material, sputter the target, and generate secondary charged and neutral products~\cite{Thuppul2020,Collins2019}. Grid impingement and the resulting secondary products can distort the extraction field, increase beam divergence, contaminate nearby hardware, and bias ground-based diagnostics \cite{Thuppul2020,Uchizono2021,Petro2022}. An energy-resolved description of the chemistry initiated by propellant--surface collisions is therefore needed to connect plume composition to component lifetime and to interpret time-of-flight and residual-gas measurements.

The prototypical propellant 1-ethyl-3-methylimidazolium tetrafluoroborate, denoted \EMIBF, is a particularly demanding atomistic test. A neutral pair of \EMIBF contains ionic, covalent, hydrogen-bonding, dispersion, and polarization interactions. During impact, the system can separate into \EMIplus\ and \BFminus, rearrange through H and F transfer, or undergo extensive cleavage of the imidazolium and tetrafluoroborate moieties \cite{Laws2026}. The relevant products span intact heavy fragments, light radicals, and volatile molecules such as HF. A useful model must therefore describe both near-equilibrium ion-pair structure and strongly nonequilibrium bond breaking over a short collision event. Bendimerad and Petro \cite{Bendimerad2022,BendimeradCorrection2024} established two classical baselines for this problem. A nonreactive hybrid model was used to map ionic dissociation, while ReaxFF enabled covalent bond breaking and pseudomass-spectrum prediction. The corrected ReaxFF study found increasing production of light fragments with impact energy and identified BF$_4$, BF$_3$, BF$_2$, F, and cation-derived products.

Laws and Petro \cite{Laws2026} subsequently used a mixed quantum/classical DFT-based molecular-dynamics framework (hereafter DFT/MD), treating the complete 24-atom projectile quantum mechanically while representing a rigid 2000-atom Au slab classically. Across 10--100 \eV, their trajectories reported a three-regime sequence: ionic dissociation at 10--20 \eV, a neutralization window at 30--40 \eV, and extensive covalent fragmentation above 50 \eV. This treatment identified HF and other neutral products that were not reported in the ReaxFF trajectory ensemble \cite{Bendimerad2022}. However, they reported that a 2~ps microcanonical trajectory required up to 60 days on 16 CPU cores with 128~GB of memory. Complementary ab initio and experimental studies likewise show that \EMIBF\ fragmentation is strongly energy dependent and that both ionic and covalent channels contribute under electrospray-relevant conditions, although their collision environments differ from neutral surface impact \cite{Sampson2026,Bell2026,Abu2025}. The reported DFT-based cost therefore limits direct sampling of the hundreds to millions of trajectories needed to propagate orientation, surface, and plume uncertainties.

Machine-learning interatomic potentials (MLIPs) offer a promising intermediate-cost approach. The message-passing atomic cluster expansion (MACE) architecture \cite{Batatia2022} combines equivariant graph representations with high-body-order message passing to approximate reference potential-energy surfaces. MACE-MP-0 \cite{Batatia2025} is a broadly trained, general-purpose foundation model that has been evaluated for bulk periodic materials, complex liquid--surface interfaces, and reactive molecular dynamics. For the present ionic-liquid system, \MACEPolar\ \cite{BatatiaPolar2026} was of particular interest because it incorporates long-range electrostatics, polarizable induction, and global charge and spin controls, all of which are expected to be relevant to the interactions and reactions of charged molecular species. The model was trained on approximately 100 million hybrid-DFT calculations from the OMol25 molecular dataset. The comparison therefore evaluates both the transferability of a general-purpose atomistic foundation model and the potential benefits of explicitly representing electrostatic and polarization effects in ionic-liquid surface-impact simulations.

This work investigates the extent to which these pretrained foundation architectures can capture selected chemical and structural outcomes of $\text{EMI-BF}_4\text{--Au}$ impacts in the absence of system-specific training. First, we evaluate the models' ability to reproduce a stable equilibrium contact ion-pair geometry. Second, we assess their performance in reproducing the energy-dependent fragmentation pathways and $\text{HF}$-forming chemistry established by mixed quantum/classical simulations. Third, we quantify their computational cost and resulting wall-time ratios relative to the DFT/MD and ReaxFF references. The resulting benchmark provides a validation map for deploying universal machine-learning potentials to scale up atomistic simulations of ionic-liquid electrospray propulsion.

\section{Methods}

\subsection{Benchmark design and foundation models}

\begin{figure}[!h]
    \centering
    \includegraphics[width=0.90\linewidth]{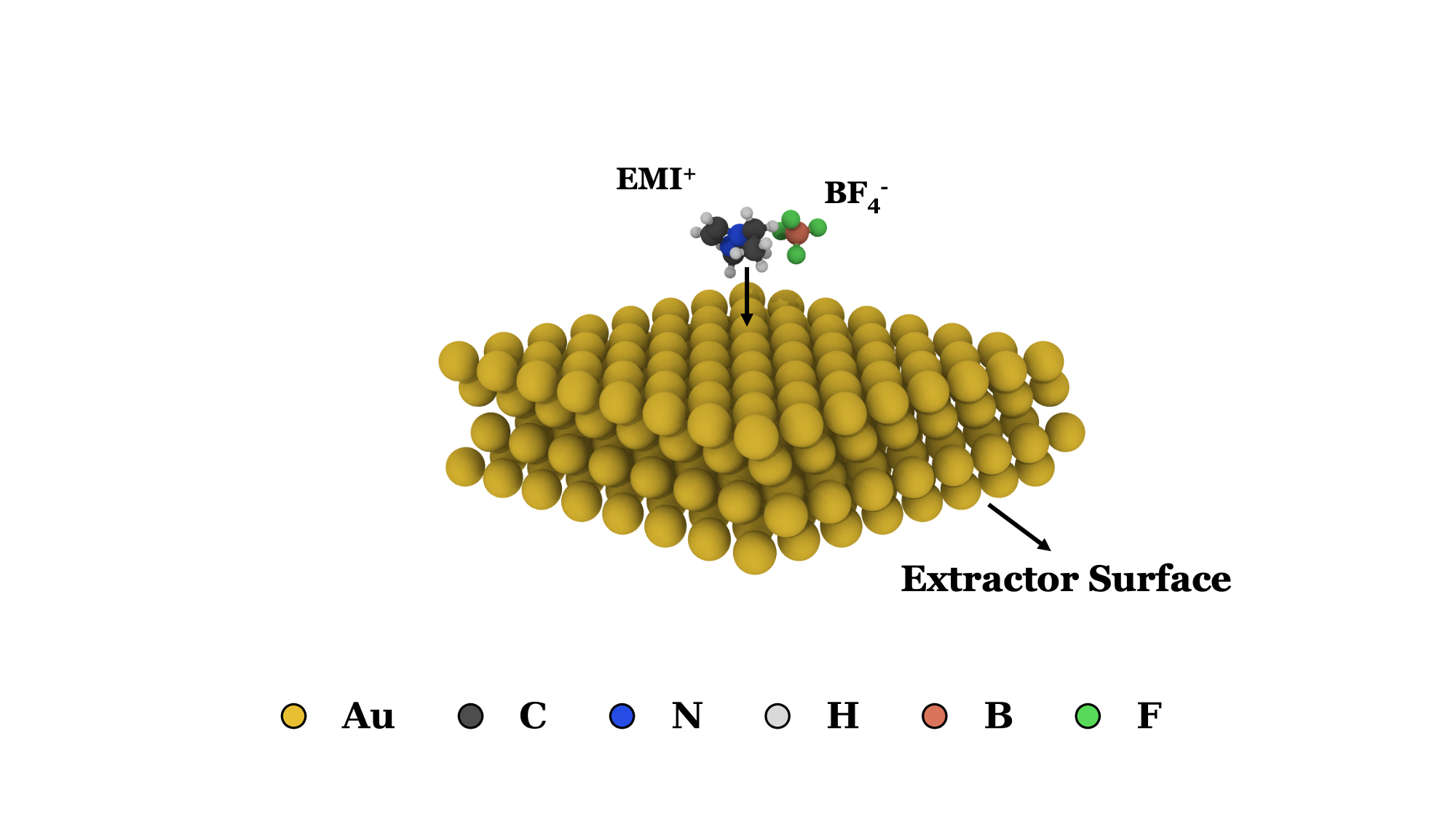}
    \caption{Simulation configuration for a neutral \EMIBF\ projectile incident on an Au extractor surface. Au, C, N, H, B, and F atoms are shown in gold, gray, blue, white, orange, and green, respectively. The MLIP trajectories follow the Au-slab geometry used by the DFT/MD reference \cite{Laws2026}. The earlier ReaxFF reference used a fixed Lennard-Jones wall rather than explicit Au atoms \cite{Bendimerad2022}.}
    \label{fig:setup}
\end{figure}

The benchmark follows the neutral-projectile surface-collision geometry of Laws and Petro \cite{Laws2026}. The incident \EMIBF\ contact ion pair contains 24 atoms, and its center of mass is initially located 20~\angstrom\ above the surface at the center of a $50\times50\times10$~\angstrom$^3$ Au slab containing 2000 atoms (Fig.~\ref{fig:setup}). The projectile is accelerated normal to the surface. Seven impact energies, 10, 20, 30, 40, 50, 75, and 100 \eV, and six orientation configurations are considered, giving 42 trajectories per model. The kinetic energy $E_{\mathrm{imp}}$ is imposed by adding the same center-of-mass velocity to every projectile atom,
\begin{equation}
v_{\mathrm{COM}}=\sqrt{\frac{2E_{\mathrm{imp}}}{m}},
\label{eq:velocity}
\end{equation}
where $m$ is the total projectile mass. The DFT/MD reference data are taken directly from Laws and Petro \cite{Laws2026}, who treated the projectile and all impact products with first-principles unrestricted Kohn--Sham DFT ($\omega$B97X-V/aug-cc-pVTZ) over a 2~ps window against a rigid, classical Au slab. The ReaxFF comparison follows the corrected results reported by Bendimerad, Laws, and Petro \cite{Bendimerad2022,BendimeradCorrection2024}.

Within this collision framework, two pretrained MACE models are evaluated. The first, $\text{MACE-MP-0}$\cite{Batatia2025}, serves as a universal materials foundation model utilizing the public \texttt{medium} checkpoint via the standard model repository \cite{Batatia2025}. The second, $\text{MACE-POLAR-1}$, augments the local MACE architecture with explicit long-range electrostatic and polarizable induction terms, trained against hybrid-DFT molecular data \cite{BatatiaPolar2026}; this implementation uses the public \texttt{polar-1-m} checkpoint. For \MACEPolar\ projectile--surface simulations, the entire system is assigned a global charge of $Q = 0$ and a spin multiplicity of 1. MACE-MP-0 has no corresponding global charge or spin inputs and is applied to the same net-neutral atomic composition. For \MACEPolar, the global constraint enforces net neutrality across the simulation domain but does not prescribe a fixed fragment topology: bonding patterns evolve during impact according to the learned potential. Because no common fragment-charge analysis is performed for all methods, the present comparison assigns product compositions from geometry but does not claim formal fragment charges.

\subsection{Geometry optimization and MD simulation setup}

  The isolated neutral contact ion pair is optimized independently with both
  MLIPs using BFGS geometry optimization with a force convergence threshold
  of \(0.03~\mathrm{eV}\,\text{\AA}^{-1}\) and a maximum of 200 optimization steps.
  Post-optimization sampling uses Maxwell--Boltzmann initialization at
  173 K, removal of net translation and rotation, and Langevin dynamics at
  173 K with a 0.5 fs timestep, a friction coefficient of
  \(0.01~\mathrm{fs^{-1}}\), 1 ps of equilibration, and 5 ps of production.
  Bond-distance distributions are obtained from
  production configurations sampled every 5 fs.

For each impact calculation, the initial EMI--BF$_4$ projectile is rotated into one of six incidence orientations and assigned the velocity from Eq.~\eqref{eq:velocity}. Both the MACE-medium and MACE-POLAR-1 trajectories use an explicit frozen Au(111) slab with the repulsive-wall implementation. Each trajectory is propagated for 4000 Velocity-Verlet steps with a 0.5~fs timestep, corresponding to 2~ps, with early termination when the maximum interatomic separation reaches 40~\AA{} to avoid pathological growth of the long-range reciprocal-space grid. The model and integration settings are held fixed across energies. Atomic configurations, fragment traces, bond traces, and model potential, kinetic, and total energies are written every 10 MD steps, i.e., every 5~fs, for post-processing.

\subsection{Fragment and spectrum analysis}

Chemical products are identified from connected components of an element-aware bond graph at the last frame of the simulation. Two atoms are considered bonded when their separation is less than \(1.3(r_i^{\mathrm{cov}}+r_j^{\mathrm{cov}})\), where \(r_i^{\mathrm{cov}}\) and \(r_j^{\mathrm{cov}}\) are tabulated covalent radii for the corresponding elements. Covalent fragmentation is assigned when the projectile contains more connected components than its initial state, and persistent bond breaking requires the relevant bond to exceed \(1.5\times\) its initial distance for five consecutive saved frames. Frames are saved every 10 MD steps with a 0.5~fs timestep, corresponding to 5~fs between saved configurations. Terminal product identities are taken from the connected components in the final saved structure. Their nominal masses are computed from standard atomic masses and rounded to the nearest atomic mass unit. At each energy and for each MACE model, the pseudomass intensity is the number of terminal fragment occurrences across the six orientations divided by the total number of terminal fragments at that energy. DFT/MD and ReaxFF spectra and occurrence data are taken from the corresponding reference datasets \cite{Laws2026,Bendimerad2022,BendimeradCorrection2024}.

\subsection{Wall-clock benchmark}

Computational cost was evaluated using the wall time for a nominal 2~ps trajectory. The DFT/MD timing was taken directly from Laws and Petro \cite{Laws2026}, who reported an upper-bound wall time of 60 days for one microcanonical trajectory using 16 CPU cores and 128~GB of memory \cite{Laws2026}. MACE-medium and MACE-POLAR-1 timings were benchmarked locally on an NVIDIA RTX 4090 GPU for 2~ps trajectories propagated with a 0.5~fs timestep. Model loading time was not included in the MLIP timing. ReaxFF timing was benchmarked locally on one Intel i9-13900K CPU thread by averaging the elapsed wall time of the 42 matched 2~ps trajectories. Because the platforms and algorithms differ, these measurements are reported as workflow wall times rather than hardware-normalized floating-point performance. For compact comparison, the workflow wall-time ratio is defined as \(R=t_{\mathrm{DFT/MD}}/t_{\mathrm{method}}\).

\section{Results and Discussion}
\label{sec:results_discussion}

\subsection{Equilibrium ion-pair structure geometry validation}

Before evaluating collision-induced fragmentation, the ability of the pretrained MACE models to describe the equilibrium EMI--BF$_4$ contact ion pair was assessed. This is an important first step because an inaccurate initial structure, particularly a weakened or excessively elongated bond, could strongly bias the subsequent dissociation pathways. The optimized geometries obtained with \MACEMedium\ and \MACEPolar\ were therefore compared with the crystallographic measurements at 173~K \cite{Choudhury2005} and previously reported DFT structure of Laws \textit{et al.}~\cite{Laws2026}. The complete comparison of anion and cation bond lengths and angles is summarized in Table~\ref{tab:complete_mace_geometry_benchmark}.

Both pretrained models produce stable and chemically reasonable contact-ion-pair geometries. 
The imidazolium-ring structure is reproduced particularly well, with cation bond-length deviations generally below 0.04~\AA\ and internal ring-angle deviations within approximately 3$^\circ$ of the experimental values \cite{Choudhury2005}.
The agreement obtained by both models is notable because they were applied directly in their pretrained form, without fine-tuning to EMI--BF$_4$, ionic liquids, or electrospray-fragmentation configurations. These results are consistent with the local covalent environment of the EMI$^+$ cation lying within the chemical domain represented by the foundation models.

The most significant differences among the methods occur within the BF$_4^-$ anion. The published DFT geometry reproduces three B--F distances accurately, with B$_1$--F$_1$, B$_1$--F$_2$, and B$_1$--F$_3$ differing from experiment by less than 0.03~\AA. However, B$_1$--F$_4$ is elongated to 1.703~\AA, compared with the crystallographic value of 1.399~\AA. Laws and Petro attributed this elongation to directional C--H$\cdots$F bonding and polarization in the isolated contact pair, whose environment differs from that of the crystal \cite{Laws2026}. The resulting isolated-pair anion is more anisotropic, with calculated F--B--F angles ranging from 94.0$^\circ$ to 123.3$^\circ$, substantially broader than the experimental range of 108.7$^\circ$--111.1$^\circ$.

However, both MACE models produce a compact, approximately tetrahedral BF$_4^-$ unit. The four optimized B--F distances lie within 1.37--1.44~\AA, while the corresponding F--B--F angles remain close to the tetrahedral experimental values. Although several individual short B--F bonds are reproduced more accurately by the published DFT calculation, the MACE models give a more compact description of the complete anion geometry without the single strongly elongated bond. Relative to the crystallographic reference, the MACE models show closer agreement for the longest B--F bond and the overall angular range, whereas DFT is closer for three short B--F bonds.

\begin{figure}[!h]
    \centering
    \includegraphics[width=\linewidth]{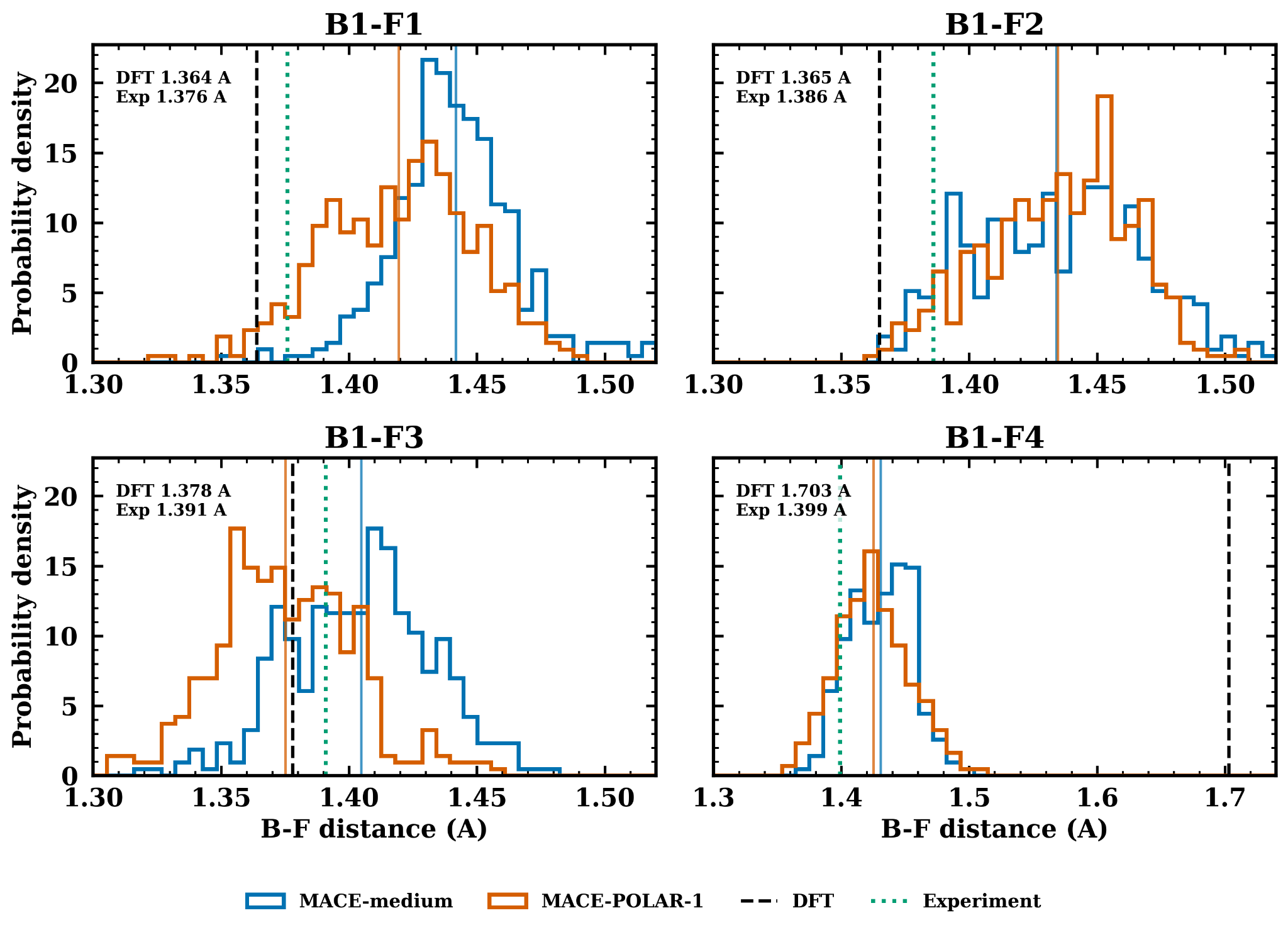}
    \caption{B--F bond-distance probability distributions for \MACEMedium\ (blue) and \MACEPolar\ (orange). Dashed black and dotted green lines denote the optimized DFT and 173~K crystallographic distances, respectively \cite{Laws2026,Choudhury2005}. Thin solid vertical lines mark the MLIP mean values. The principal structural distinction is that neither MLIP retains the longer isolated-pair DFT B$_1$--F$_4$ bond; DFT remains closer to experiment for several individual short bonds.}
    \label{fig:bond_dist}
\end{figure}

To test finite-temperature stability, we further performed canonical (NVT) molecular-dynamics simulations at 173~K with both MACE models. Fig.~\ref{fig:bond_dist} shows the resulting B--F bond-distance distributions from the 5~ps production sampling following 1~ps of equilibration. Across the sampled trajectories, all four bonds remain within the bonded region and fluctuate around compact equilibrium distances, with no persistent population corresponding to the strongly elongated B$_1$--F$_4$ bond reported in the published DFT geometry. \MACEPolar\ yields a slightly shorter B$_1$--F$_3$ distribution, while the remaining bond distributions predicted by the two models are broadly similar, indicating a consistent description of the anion structure under thermal sampling.

The crystallographic reference \cite{Choudhury2005} corresponds to a condensed-phase environment in which neighboring ions influence the anion geometry, whereas the computational structures represent an isolated contact ion pair. The comparison should therefore be interpreted as a structural validation rather than as a general ranking of MACE and DFT accuracy. Nevertheless, its implication for the present collision study is direct. Both pretrained models generate a stable initial ion pair with realistic cation geometry and a compact BF$_4^-$ anion, ensuring that the subsequent fragmentation trajectories begin without the longer isolated-pair DFT B--F bond. The reproduction of this initial structure supports evaluation of pretrained MACE foundation models for ionic-liquid collision chemistry without system-specific retraining.

\begin{table}[htbp]
\centering
\captionsetup{labelformat=empty}
\caption{TABLE I. Comparison of DFT optimized EMI--BF$_4$ geometry metrics
with experimental crystallography measurements at 173~K and with MACE-medium
and MACE-POLAR-1 optimized geometries. Bond lengths are in \AA;
angles are in degrees.}
\label{tab:complete_mace_geometry_benchmark}
\small
\begin{tabular}{lcccc}
\toprule
Measurement & DFT/MD \cite{Laws2026} & Experimental \cite{Choudhury2005} & MACE-medium  & MACE-POLAR-1 \\
\midrule
\multicolumn{5}{l}{\textbf{Anion bond lengths (\AA)}} \\
B$_1$--F$_1$ & 1.364 & 1.376 & 1.438 & 1.433 \\
B$_1$--F$_2$ & 1.365 & 1.386 & 1.430 & 1.419 \\
B$_1$--F$_3$ & 1.378 & 1.391 & 1.402 & 1.371 \\
B$_1$--F$_4$ & 1.703 & 1.399 & 1.430 & 1.420 \\
\addlinespace
\multicolumn{5}{l}{\textbf{External anion angles (deg)}} \\
F$_1$--B$_1$--F$_2$ & 94.0 & 108.7 & 108.7 & 106.8 \\
F$_1$--B$_1$--F$_3$ & 96.0 & 108.8 & 109.8 & 111.2 \\
F$_1$--B$_1$--F$_4$ & 96.8 & 109.0 & 108.6 & 106.8 \\
F$_2$--B$_1$--F$_3$ & 113.9 & 109.5 & 110.5 & 112.0 \\
F$_2$--B$_1$--F$_4$ & 120.0 & 109.7 & 109.1 & 107.8 \\
F$_3$--B$_1$--F$_4$ & 123.3 & 111.1 & 110.2 & 111.9 \\
\addlinespace
\multicolumn{5}{l}{\textbf{Cation bond lengths (\AA)}} \\
N$_1$--C$_2$ & 1.292 & 1.330 & 1.341 & 1.327 \\
C$_2$--N$_3$ & 1.305 & 1.335 & 1.342 & 1.328 \\
N$_3$--C$_4$ & 1.346 & 1.361 & 1.392 & 1.378 \\
C$_4$--C$_5$ & 1.436 & 1.384 & 1.356 & 1.354 \\
C$_5$--N$_1$ & 1.452 & 1.390 & 1.383 & 1.377 \\
\addlinespace
\multicolumn{5}{l}{\textbf{Internal cation angles (deg)}} \\
C$_5$--N$_1$--C$_2$ & 104.8 & 106.3 & 108.6 & 108.9 \\
N$_1$--C$_2$--N$_3$ & 106.5 & 107.8 & 108.5 & 108.5 \\
C$_2$--N$_3$--C$_4$ & 109.2 & 107.8 & 108.4 & 108.8 \\
N$_3$--C$_4$--C$_5$ & 109.3 & 108.9 & 107.0 & 106.9 \\
C$_4$--C$_5$--N$_1$ & 110.0 & 109.3 & 107.5 & 106.9 \\
\bottomrule
\end{tabular}
\end{table}

\subsection{Impact-induced fragmentation pathways and product branching}

\begin{figure}[!h]
    \centering
    \begin{subfigure}[t]{0.32\linewidth}
        \centering
        \includegraphics[width=\linewidth]{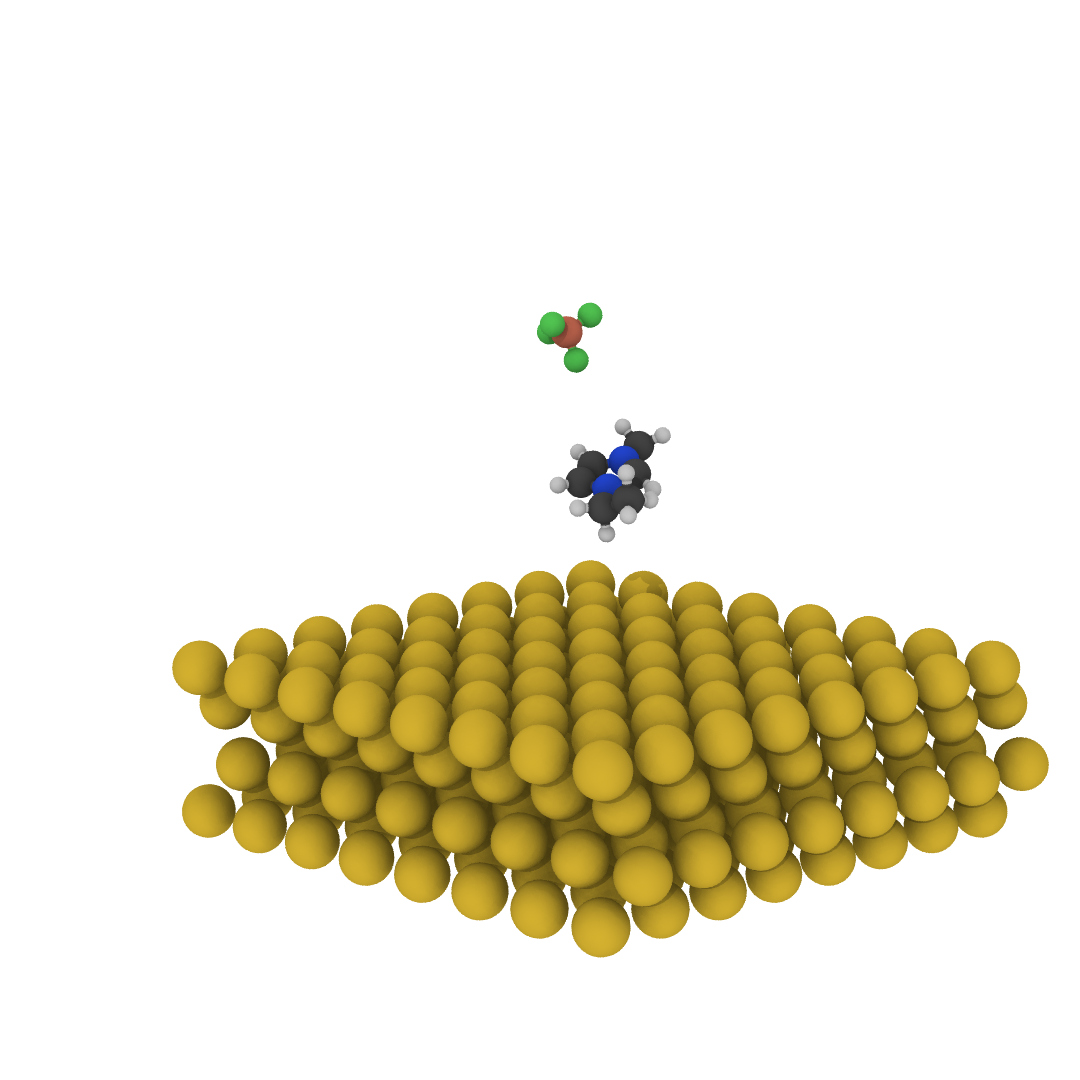}
        \caption{Ionic dissociation}
    \end{subfigure}\hfill
    \begin{subfigure}[t]{0.32\linewidth}
        \centering
        \includegraphics[width=\linewidth]{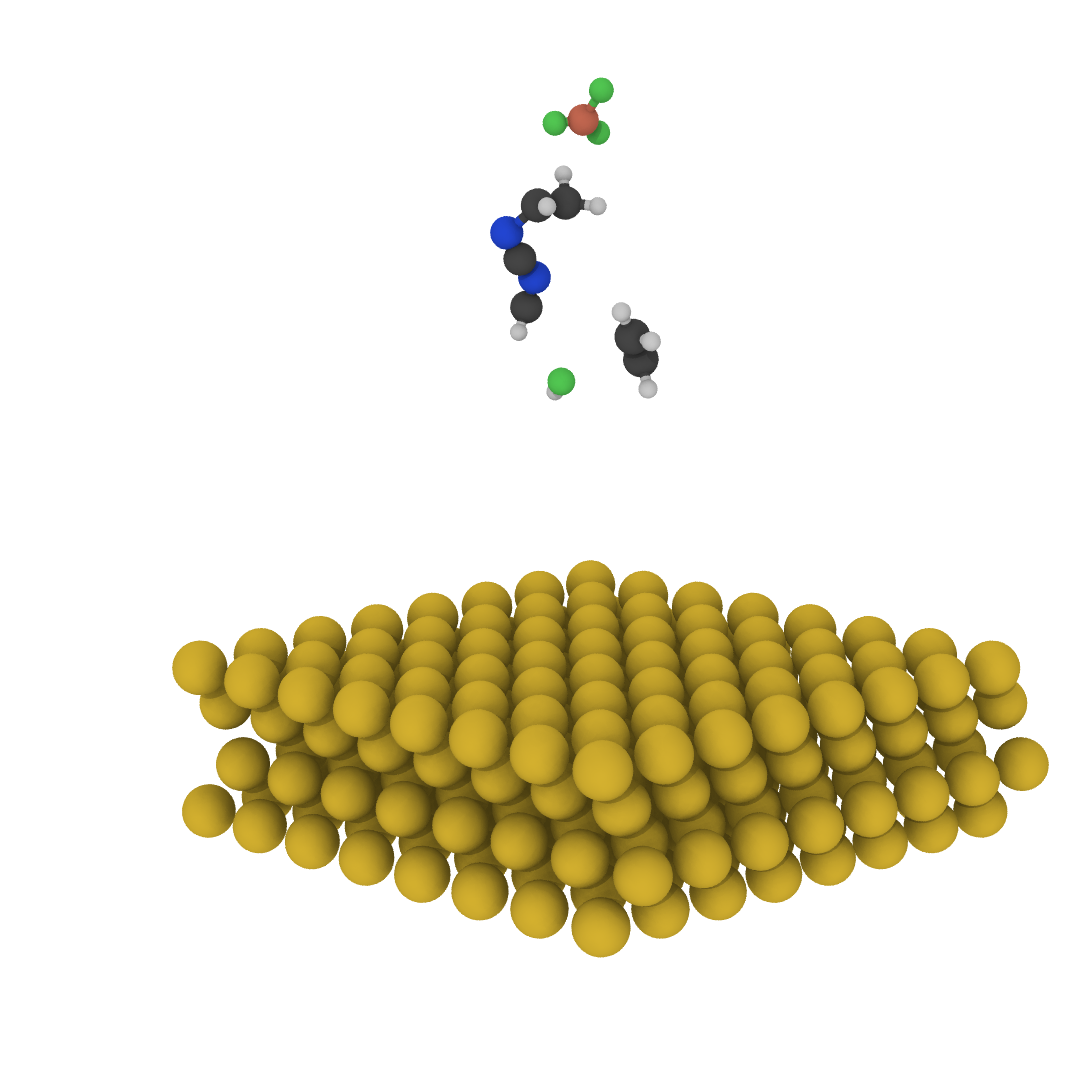}
        \caption{HF-forming rearrangement}
    \end{subfigure}\hfill
    \begin{subfigure}[t]{0.32\linewidth}
        \centering
        \includegraphics[width=\linewidth]{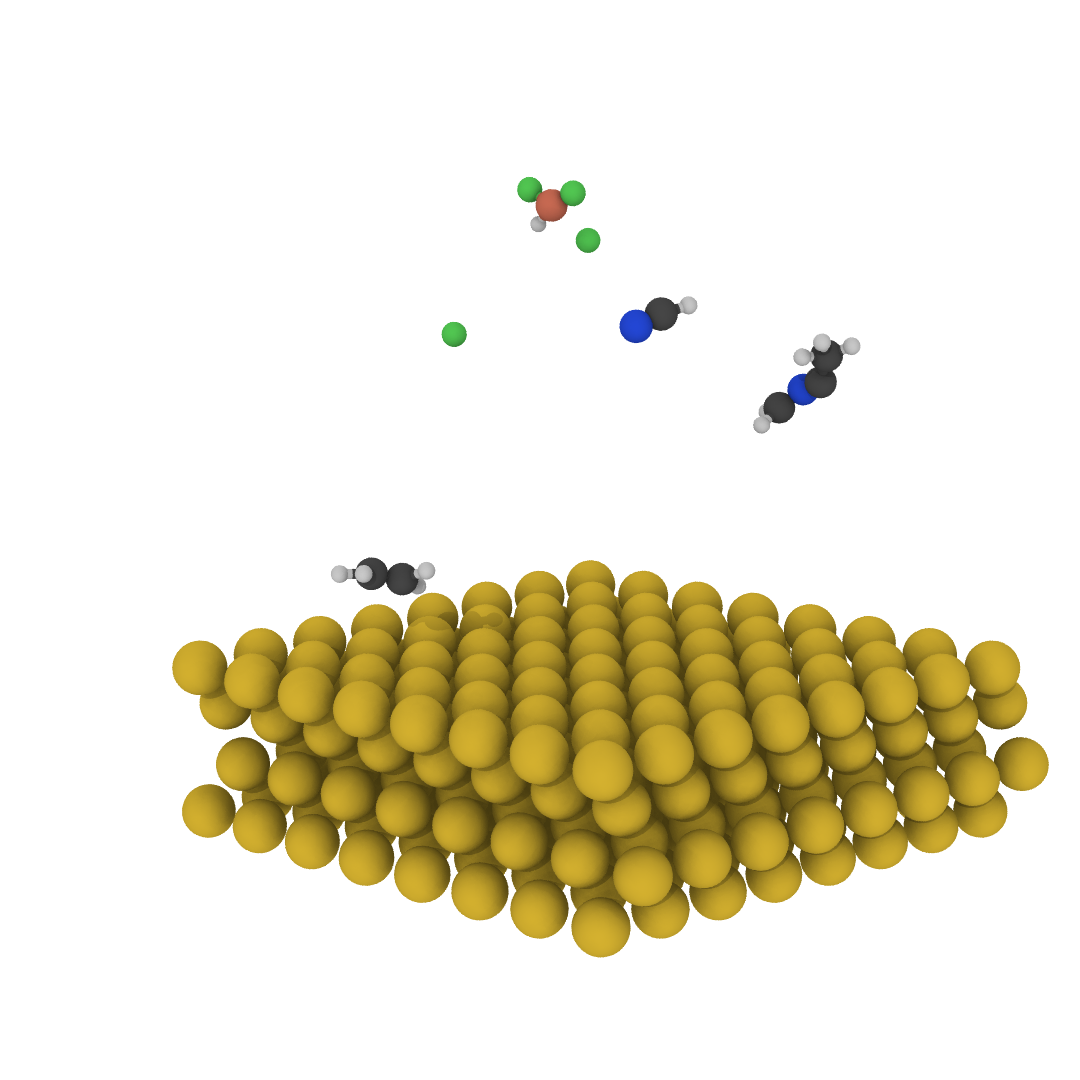}
        \caption{Covalent fragmentation}
    \end{subfigure}
    \caption{Representative post-impact configurations from \MACEPolar. (a) Separation of intact ionic constituents; (b) an intermediate-energy pathway producing BF$_3$, HF, and a cation-derived backbone; and (c) extensive high-energy covalent fragmentation. Colors follow Fig.~\ref{fig:setup}.}
    \label{fig:pathways}
\end{figure}

The optimized geometry is then used as the initial configuration for the surface-impact simulations. At each impact energy, the pseudomass spectra and species occurrences are aggregated over the six incidence orientations. Fig.~\ref{fig:pathways} presents representative \MACEPolar\ configurations selected from this ensemble to illustrate the same three broad regimes identified in the DFT/MD study. At low energy (10~eV), four of the six trajectories show separation of the largely intact \EMIplus\ and \BFminus\ constituents with limited covalent damage, whereas the remaining two exhibit moderate fragmentation accompanied by HF formation. At intermediate energy (20-50 ev), hydrogen transfer from the cation to fluorine produces HF, while the boron-containing fragment approaches BF$_3$ stoichiometry and the remaining cation-derived backbone undergoes neutralization-like rearrangement. At high energy (> 75 eV), multiple covalent bonds rupture, generating light B/F fragments, N/C-containing fragments, and hydrocarbons.

The appearance of these pathways in zero-shot simulations shows that the pretrained model is not restricted to preserving the input bonding topology. The sampled configurations include stretched ionic contacts, strongly compressed projectile--surface geometries, simultaneous bond breaking and formation, and radical-like products. Their energy ordering is consistent with the DFT/MD sequence: intact ionic dissociation at low energy, H/F transfer and neutralization-like chemistry at intermediate energy, and widespread covalent cleavage at high energy.

\begin{figure}[!h]
    \centering
    \includegraphics[width=\linewidth]{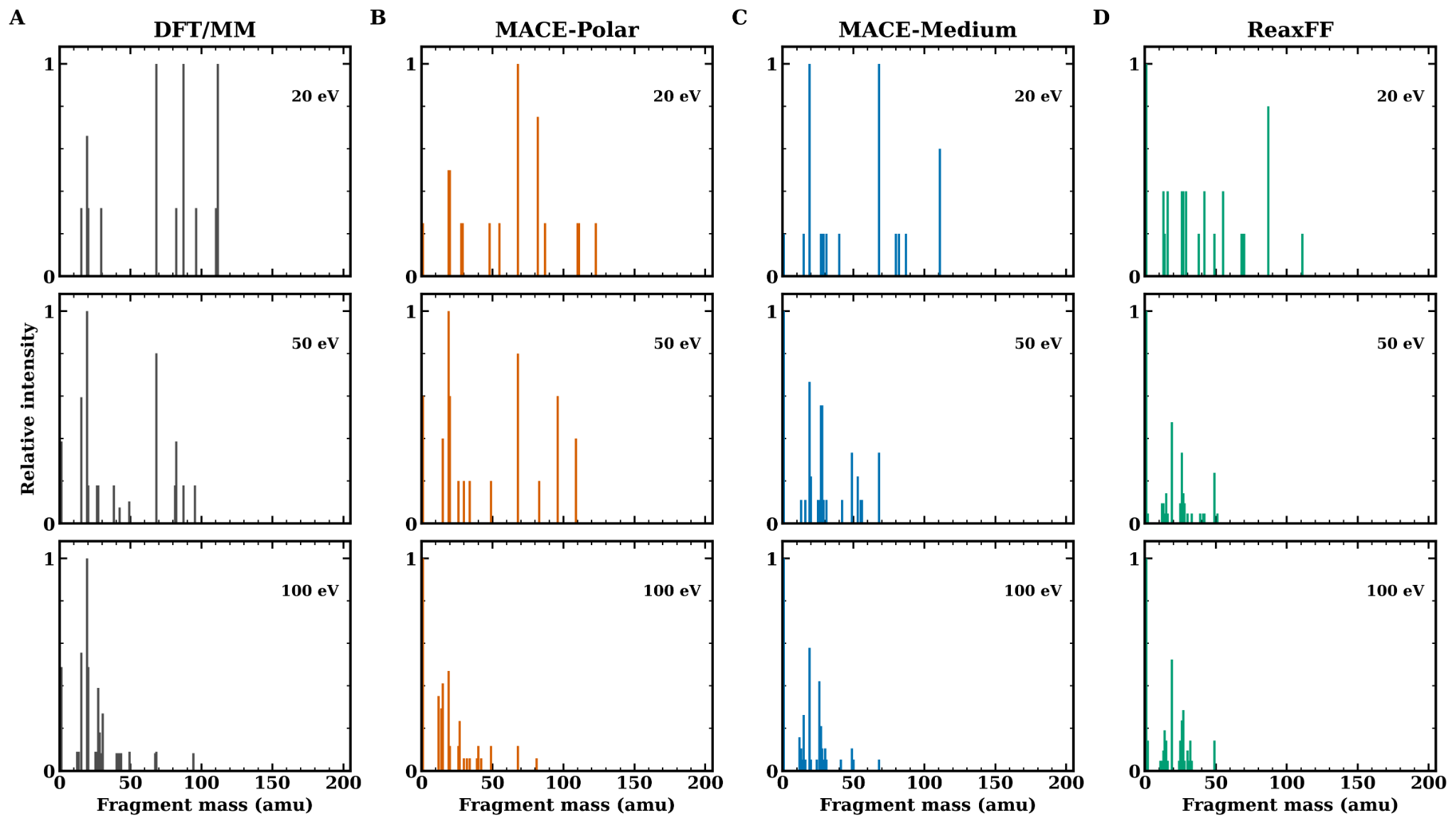}
    \caption{Normalized pseudomass spectra at 20, 50, and 100 \eV\ from (A) DFT/MD, (B) \MACEPolar, (C) \MACEMedium, and (D) ReaxFF. The MLIP spectra aggregate terminal products over six incidence orientations. Each row is normalized within its method and energy, so peak heights compare relative composition within a panel rather than absolute product yields across methods. DFT/MD and ReaxFF reference data follow Refs.~\cite{Laws2026,Bendimerad2022,BendimeradCorrection2024}.}
    \label{fig:mass_spectra}
\end{figure}

Fig. ~\ref{fig:mass_spectra} compares normalized pseudomass spectra at 20, 50, and 100 \eV. Each MLIP spectrum pools the terminal fragments from all six incidence orientations at the corresponding energy before normalization. All four methods show the expected shift from relatively heavy, few-body products at low energy toward many light fragments at high energy. At 20 \eV, DFT/MD retains strong high-mass channels associated with ionic dissociation. \MACEPolar\ likewise maintains a broad distribution containing intermediate and heavy fragments, whereas \MACEMedium\ and ReaxFF place more intensity in light products. At 50 \eV, \MACEPolar\ preserves a DFT-like mixture of light, intermediate, and heavy products. By 100 \eV, the spectra from every reactive method are dominated by fragments below approximately 60~amu, consistent with extensive covalent cleavage.

\begin{figure}[!h]
    \centering
    \includegraphics[width=\linewidth]{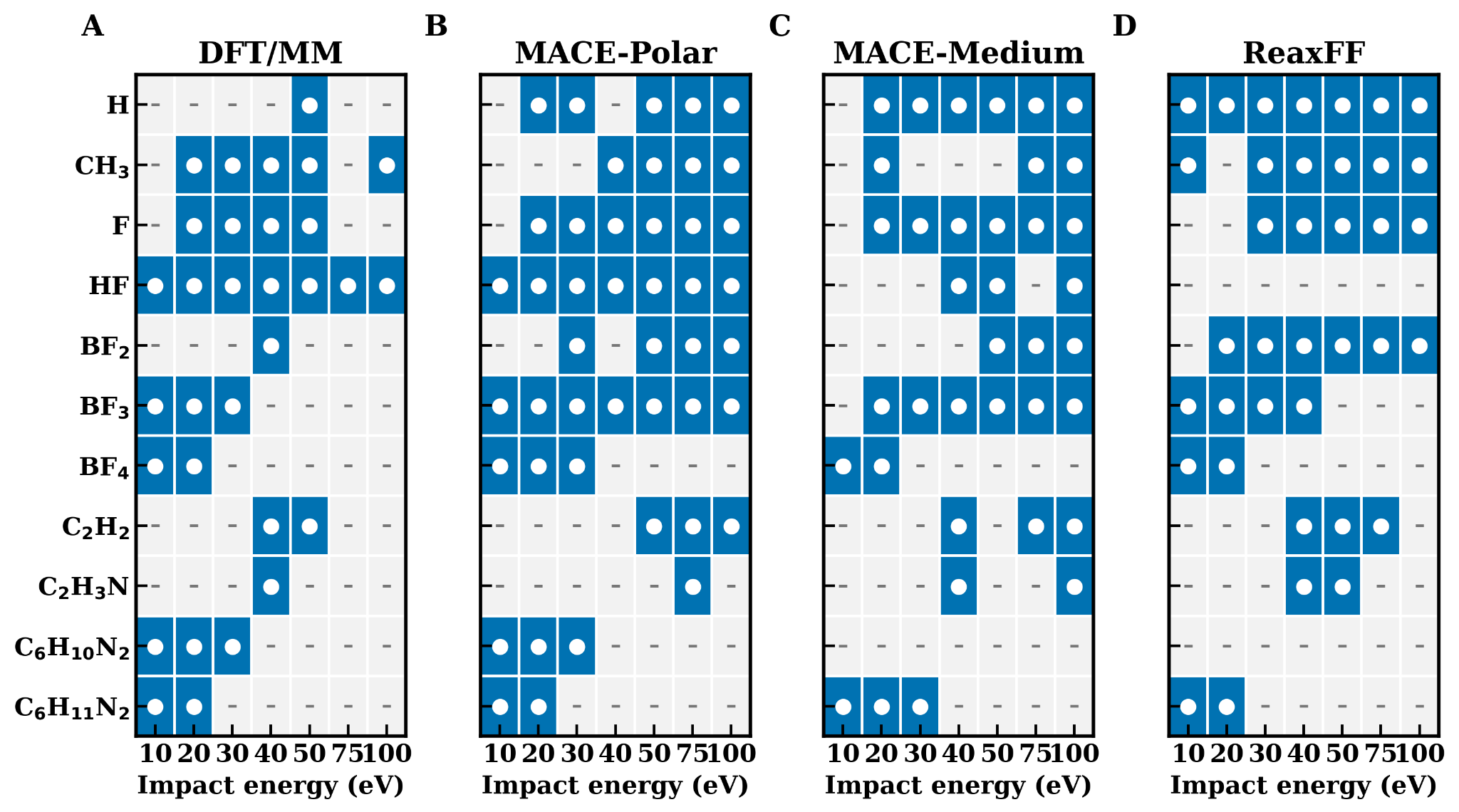}
    \caption{Occurrence of 11 diagnostic products over 10--100 \eV\ for (A) DFT/MD, (B) \MACEPolar, (C) \MACEMedium, and (D) ReaxFF. Blue cells indicate that a species was identified in at least one trajectory at that energy; gray cells indicate non-observation. Presence/absence does not encode yield or formal charge.}
    \label{fig:species_map}
\end{figure}

The mass spectra confirm that both pretrained MLIPs access reactive potential-energy landscapes. However, products with similar nominal masses may arise from different bond-breaking and bond-forming pathways. The species--energy occurrence map in Fig.~\ref{fig:species_map} therefore provides a more chemically specific comparison. Relative to DFT/MD, \MACEPolar\ recovers 26 of 31 positive species--energy cells, corresponding to 83.9\% positive-cell coverage. These common cells include diagnostic BF$_3$, BF$_2$, F, HF, and cation-derived products also found in the DFT/MD trajectories. The \MACEMedium\ occurrence pattern is more similar to the ReaxFF reference overall: their maps agree in 55 of 77 cells (71.4\%), while \MACEMedium\ nevertheless contains several DFT/MD products that were not observed in the ReaxFF trajectories.

 HF formation provides the clearest species-specific distinction among the methods. Capturing the formation of HF is important because Laws and Petro identify the reaction EMI--BF$_4\rightarrow$ C$_6$H$_{10}$N$_2$ + BF$_3$ + HF as a marker of neutralization chemistry \cite{Laws2026}. As a neutral product, HF is not recorded by charged-particle time-of-flight or retarding-potential measurements, but it is accessible to complementary residual-gas analysis \cite{Laws2026,Shaik2024}. DFT/MD identifies HF at every sampled energy, consistent with hydrogen transfer to fluorine and with experimental observations of volatile collision products \cite{Laws2026,Shaik2024}. \MACEPolar\ likewise forms HF at all seven energies, whereas \MACEMedium\ forms HF at 40, 50, and 100~\eV. No HF is reported in the corrected ReaxFF reference trajectories. Thus, although the overall \MACEMedium\ occurrence pattern is similar to ReaxFF, the pretrained model samples an HF-forming pathway not reported in that ReaxFF ensemble. Bendimerad and Petro \cite{Bendimerad2022,BendimeradCorrection2024} reported free H and noted that their simulation time and length scales did not resolve its subsequent fate. Conversely, occurrence of HF at every sampled energy in \MACEPolar\ may reflect either increased sensitivity to hydrogen-transfer chemistry or an overly reactive description. Product occurrence alone cannot distinguish these possibilities.

The contrast between the two MLIPs may reflect their different architectures and training domains. \MACEMedium\ was trained primarily on periodic PBE materials configurations and represents interactions through finite-range equivariant message passing \cite{Batatia2025}. It can describe bond rearrangement and interfacial chemistry, but it does not explicitly impose the long-range electrostatic response of a separating ion pair. \MACEPolar\ was instead developed for molecular chemistry with explicit long-range electrostatics, iterative polarization, and global charge and spin control, using a hybrid-DFT molecular training domain \cite{BatatiaPolar2026}. These features may be relevant to the competition among ionic separation, charge redistribution, hydrogen transfer, and covalent fragmentation in \EMIBF. 

\subsection{Computational performance and application regimes}

\begin{figure}[t]
    \centering
    \includegraphics[width=0.90\linewidth]{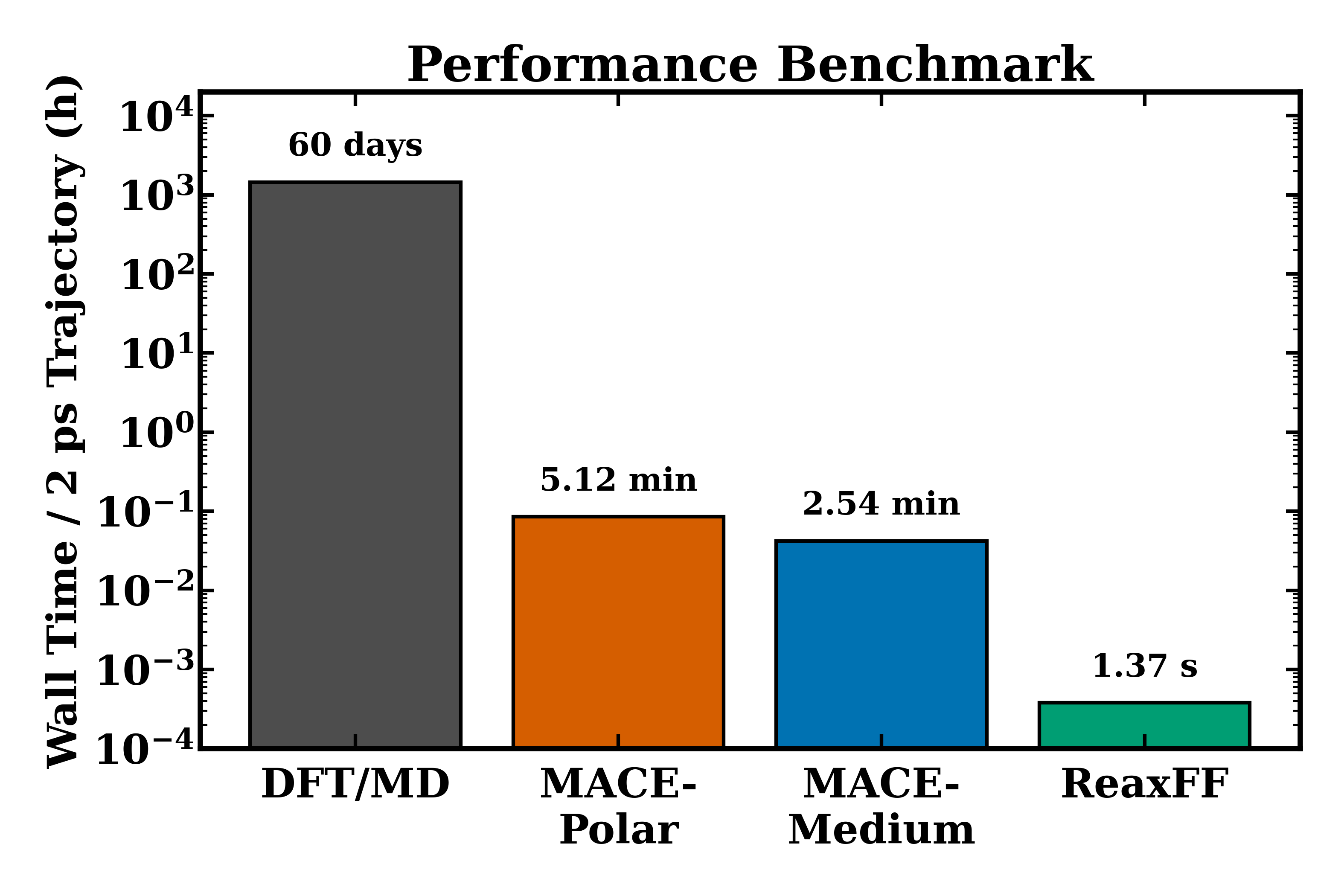}
    \caption{Wall time per 2~ps trajectory for DFT/MD, \MACEPolar, \MACEMedium, and ReaxFF. The logarithmic scale emphasizes the intermediate cost of the MLIPs. Timings are workflow specific and are not normalized for hardware.}
    \label{fig:performance}
\end{figure}

The computational cost spans nearly seven orders of magnitude across the four approaches (Fig.~\ref{fig:performance} and Table~\ref{tab:performance}). A 2~ps DFT/MD trajectory requires up to 1440~h \cite{Laws2026}, whereas the same nominal trajectory length requires only 0.0853~h (5.12~min) with \MACEPolar\ and 0.0423~h (2.54~min) with \MACEMedium. These timings correspond to workflow wall-time ratios of $1.69\times10^4$ and $3.40\times10^4$ relative to the reported DFT/MD upper bound, respectively. ReaxFF remains the least expensive method, requiring only 1.37~s per trajectory and running approximately 111 times faster than \MACEMedium\ and 224 times faster than \MACEPolar.

\begin{table}[t]
\centering
\caption{Observed wall time and ratio to the reported DFT/MD upper bound per 2 ps trajectory.}
\label{tab:performance}
\begin{tabular}{lrr}
\toprule
Method & Wall time & Wall-time ratio vs DFT/MD \\
\midrule
DFT/MD       & 60 d        & 1 \\
\MACEPolar   & 5.12 min    & $1.69\times10^4$ \\
\MACEMedium  & 2.54 min    & $3.40\times10^4$ \\
ReaxFF       & 1.37 s      & $3.78\times10^6$ \\
\bottomrule
\end{tabular}
\end{table}

This cost hierarchy defines complementary application regimes rather than a simple ranking of methods. DFT/MD remains appropriate for reference calculations on selected structures, reaction pathways, and electronically complex events, but its cost restricts the number of trajectories that can be sampled. ReaxFF enables much larger ensembles when an appropriate parameterization has already been validated, although HF was not reported under the sampled protocol in the corrected reference study. The pretrained MLIPs occupy an intermediate regime: they reduce the cost of DFT/MD by more than four orders of magnitude while sampling HF-forming pathways and, for \MACEPolar, showing greater DFT/MD-positive-cell coverage than the ReaxFF reference. This balance is most evident in the positive-cell coverage of \MACEPolar\ and the ability of both MACE models to access HF-forming pathways.

The resulting throughput changes the type of collision study that can be performed. Rather than representing each impact energy using only a few trajectories or a prescribed fragmentation rule, MLIP simulations can sample distributions over projectile orientation, internal temperature, incidence angle, and surface configuration. Such ensembles could provide statistically informed product distributions for higher-scale particle-in-cell and plume-transport calculations, including predictions of whether fragments escape through an aperture, return to an extractor grid, deposit on nearby surfaces, or contribute to diagnostic backgrounds. The reported timings represent realized workflow throughput rather than hardware-normalized algorithmic performance; nevertheless, a reduction of more than four orders of magnitude relative to DFT/MD indicates that pretrained MLIPs can make chemically resolved collision ensembles computationally practical and provide a possible pathway toward increasingly quantitative plume, contamination, and component-level predictions.

\section{Conclusions}
Two pretrained MACE potentials were benchmarked for neutral \EMIBF\ fragmentation on a model Au extractor surface under electrospray-relevant impact energies of 10--100~\eV. Both models successfully produced compact EMI--BF$_4$ geometries consistent with crystallographic measurements \cite{Choudhury2005}, as well as energy-dependent ionic dissociation, HF-forming rearrangements, and high-energy covalent fragmentation \cite{Laws2026}. Notably, \MACEPolar\ achieved an 83.9\% positive-cell overlap with the reference DFT/MD occurrence map. Meanwhile, \MACEMedium\ aligned more closely with the ReaxFF baseline, yet uniquely captured HF-formation pathways missed by classical reactive force fields. Both MLIP workflows delivered these chemically reactive trajectories at wall times approximately four orders of magnitude lower than DFT/MD, occupying a high-throughput, intermediate-cost regime between first-principles methods and empirical force fields.

These results provide evidence that the pretrained potentials can sample selected configurations beyond their equilibrium validation regime, accurately capturing non-equilibrium bond breaking and formation during high-energy collisions. Looking forward, these findings lay the groundwork for an efficient, multi-fidelity simulation paradigm: fast empirical methods can rapidly explore vast collision configuration spaces, while targeted fine-tuning focuses computationally intensive DFT calculations preferentially on high-uncertainty impact frames—such as active B--F breaking, H-transfer coordinates, or extreme surface-contact states. Model validation should progress systematically from equilibrium structures and dissociation curves to short DFT/MD trajectory segments, and finally to withheld impact energies and orientations. Taken together, these structural, occurrence-map, and cost benchmarks establish a practical basis for developing an electrospray-specific foundation model, with \MACEPolar\ serving as a promising pre-conditioned architecture for fine-tuning.

While the present benchmark represents an early-stage, qualitative campaign across six orientations per energy, it establishes an initial validation map for deploying universal MLIPs to space propulsion problems. Future developments—including dynamic and polarizable metal substrates, explicit charge-transfer models, expanded orientation ensembles, and targeted fine-tuning—will further refine quantitative branching ratios and yield predictions. Ultimately, this work demonstrates that foundation-model MLIPs offer an effective, scalable bridge between first-principles quantum chemistry and device-scale atomistic modeling of ionic-liquid electrospray thrusters.

\section*{Funding}

This work was supported by the Georgia Institute of Technology Aerospace Engineering Postdoctoral Fellowship and Astrobiology Postdoctoral Fellowship. The authors also acknowledge the Partnership for an Advanced Computing Environment (PACE) at Georgia Tech for providing computational resources and technical support.

\section*{CRediT authorship contribution statement}
\CorrespondingAuthor: Conceptualization, Methodology, Software, Validation, Formal analysis, Investigation, Data curation, Visualization, Writing -- original draft, Writing -- review and editing. 

\section*{Declaration of competing interest}
The authors declare that they have no known competing financial interests or personal relationships that could have appeared to influence the work reported in this paper.

\section*{Data availability}
The trajectory data, fragment tables, and analysis scripts supporting this study will be made publicly available through the project GitHub repository at \url{https://github.com/ziyuhuang652/ML_IonicLiquid_Frag}.

\bibliographystyle{elsarticle-num}
\bibliography{references}

\end{document}